\documentclass[12pt,a4paper,final]{iopart}
\usepackage{graphicx}

\usepackage{iopams}  
\usepackage[breaklinks=true,colorlinks=true,linkcolor=blue,urlcolor=blue,citecolor=blue]{hyperref}
\usepackage[utf8]{inputenc}
\usepackage[english]{babel}
\expandafter\let\csname equation*\endcsname\relax
\expandafter\let\csname endequation*\endcsname\relax
 \usepackage{amsmath}
\usepackage[x11names]{xcolor}

\begin{document}

\title[]{Optical properties of magnetized transient low-pressure plasma}
\author{Roman Bergert, Slobodan Mitic}
\address{Institute of Experimental Physics I, Justus-Liebig-University Giessen, Heinrich-Buff-Ring 16,
D-35392 Giessen, Germany}
\ead{Roman.Bergert@physik.uni-giessen.de}

\begin{abstract}
A plasma under the influence of an external magnetic field changes the optical properties due to the Zeeman splitting of the energy levels. This splitting degenerates an initial single spectral line into a system of spectral lines with different transition frequencies defined by the electronic structure of the energy levels. Newly created magnetic sub-levels redefine the spectral profile of the line emission and therefore radiation transport mechanism in optically thick plasma.

Self-absorption which defines the excited state-densities is an important mechanism and can be used with other methods to describe the state densities for an optically thick plasma. This method is an established tool to retrieve state-density and plasma parameters. To measure each magnetic sub-level density of argon $\rm 1s_4$ and $\rm 1s_5$ (in Paschen's notation) a tunable diode laser absorption spectroscopy (TDLAS) was used. Based on reconstructed excited state densities, the self-absorption coefficient was calculated for individual magnetic sub-levels. A decrease in self-absorption with an external magnetic field was noticed indicating a higher transparency of the plasma. Furthermore a polarization dependent self-absorption was found. The presented results can help to model optical properties and interpret the absorption of a low-pressure optically thick magnetized plasma.
\end{abstract}


\maketitle

\section{Introduction}
The diagnostic of low-pressure plasma influenced by an external magnetic field $B$ is still a challenging task. The external field changes the plasma properties by influencing the movement of the charged particle by the Lorentz force. The velocity distribution of the particles, especially from the electrons, are no longer well-described by a Maxwell-Boltzmann distribution resulting in anisotropy of the electron energy distribution function (EEDF) \cite{iwamae2001polarization,fujimoto1996atomic}. Another important effect which is induced by the magnetic field is the Zeeman effect \cite{zeeman1897xxxii}. 
The effect describes the splitting of an energy level with total angular momentum quantum number $J$ into $2J+1$ magnetic sub-levels in the presence of a magnetic field resulting in splitting of an emission line into the system of sub-transitions symmetrically redistributed around initial unshifted line center. As a result each spectral transition will have to be rewritten as a sum of optically allowed transitions between corresponding upper and lower magnetic sub-levels \cite{taylor2017zeeman}. 
The degeneracy of the levels vanishes with higher external magnetic fields so that an individual description of transitions between magnetic sub-levels is necessary \cite{fujimoto2008plasma}. In optically thick plasmas density distribution of $\rm 1s$ sub-states should be accounted for correct description of the self-absorption effect and thus the light transport.

A population-alignment collisional-radiative model is proposed by T. Fujimoto \cite{fujimoto1997plasma} describing the magnetic sub-level densities accounting for different collisional and radiative interactions between them. The model is used to describe polarization anisotropy in optical emission spectra \cite{iwamae2005anisotropic} mostly related to high electron energies, neglecting self-absorption and in the absence of a magnetic field. Self-absorption is usually not accounted for because most studies deal with optically thin conditions as fusion plasmas at low pressures \cite{fujimoto2008plasma,fujimoto1988new} or electron beam-plasma conditions \cite{fujimoto1997plasma,beiersdorfer1997polarization}. In general, there is a lack of measurement and interpretation for argon as working gas while plenty of observations were done for helium \cite{fujimoto2008plasma}. While most of the analysis is done on helium plasmas, where collisional excitation cross-sections between magnetic sub-levels are available, literature describing such magnetic sub-state interactions in argon is rare.  

The work of C. Csambal et al \cite{csambal1997interpretation} provides an interpretation of the laser absorption profiles for a cylindrical magnetron discharge based on a frequency integrated absorption coefficient over the whole absorption structure and not distinguishing between sub-states with different magnetic quantum numbers $m$. Several recent studies were done using laser-induced fluorescence \cite{thompson2018laser} and high resolution optical emission spectroscopy \cite{taylor2017zeeman} to evaluate effects of a magnetic field on transition profiles in argon.   
However evaluation of the densities of individual magnetic sub-levels and their impact on optical properties were not discussed.

This work focuses on evaluation of laser absorption spectra in magnetized argon plasma conditions in order to estimate magnetic sub-level densities of targeted $\rm 1s$ states. Laser absorption spectroscopy (LAS) was done targeting meatastable $\rm 1s_5$ and resonant $\rm 1s_4$ magnetic sub-level states by inducing transitions to the $\rm 2p_8$ level in argon corresponding to 801.48 nm and 842.47 nm emission lines respectively. This examples are describing transitions between levels with principal total angular momentum quantum numbers $J=(2\rightarrow 2)$ and $J=(2\rightarrow 1)$ and can be further used to evaluate emission originating from other excited levels with $J=2$ ($\rm 2p_3$ and $\rm 2p_6$ states) in argon. Measured absorption structure was reconstructed by identifying individual transitions between magnetic sub-levels clearly separated due to a strong external magnetic field of 0.3 T. Individually reconstructed sub-transition profiles were correlated to the magnetic sub-level densities of targeted $\rm 1s$ states by recalculating branching of Einstein coefficient for each individual sub-transition. As a result three magnetic sub-levels ($m=-1,0,1$) of $\rm 1s_4$ and 5 magnetic sub-levels ($m=-2,-1,...,2$) of $\rm 1s_5$ states were identified and densities of each sub-state were evaluated. Evaluated magnetic sub-level densities were further used to analyze and discuss self-absorption properties of individual sub-transition and resulting effects on optical emission spectra. It was found that splitting of energetic levels in the presence of magnetic field will effect the self-absorption properties by increasing transparency of the system.  
The measurements were done on an example of low pressure transient plasma conditions, previously investigated in \cite{kaupe2018phase}, where phase resolved $\rm 1s$ magnetic sub-level densities were evaluated for wider electron temperature and density range in the presence of an external magnetic field.

\section{Methods}

Correct interpretation of line emission (absorption) profiles in magnetized plasma conditions is based on line splitting formalism explained by Zeeman effect and its further implications in description of a self-absorption effect in optically thick conditions. Direct detection of individual magnetic sub-level absorption profiles was done by scanning over a broad frequency range by a tunable diode laser.

\subsection{Anomalous Zeeman effect}
\begin{figure}[h]
	\centering
	\includegraphics[width=0.5\linewidth]{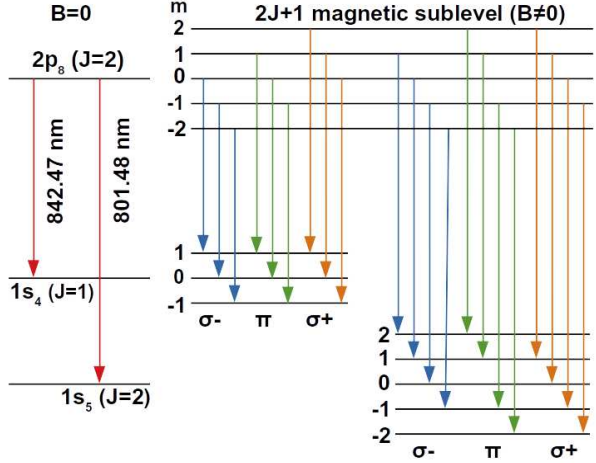}
	\caption{Representation of the energy levels for B=0 and B$\neq$0 of the used transitions where $\pi$ represents linear polarized light and $\sigma$ circularly polarized light.}
	\label{fig:schema-split}
\end{figure}

 Zeeman splitting of energy levels in a presence of magnetic field leads to systematic energy shift of degenerated sub-levels according to their magnetic quantum numbers $m$. Such effect was experimentally observed by laser induced optical transitions from $\rm 1s_5$ and $\rm 1s_4$ to $\rm 2p_8$ energy level corresponding to wavelength of 801.48 nm and 842.47 nm respectively as sketched in figure \ref{fig:schema-split}. With total angular momentum quantum number $J$ described by Russell-Saunders coupling $J=L+S$ \cite{herzberg1944atomic} the $\rm 2p_8$ level would have $J=2$ value while $\rm 1s_5$ and $\rm 1s_4$ would result in $J=2$ and $J=1$ respectively. With each level splitting on $2J+1$ magnetic sub-levels (defined by magnetic quantum numbers $m=-J,-J+1,...,J$) and accounting for selection rules, a single transition (in non magnetized case) would degenerate into the system of 9 for 842.47 nm and 12 lines for 801.48 nm transitions in a magnetic filed as shown in the right part of the figure \ref{fig:schema-split}.   

With the knowledge of the quantum numbers and the Land\'{e} factors $g_{J_{i,j}}$, the frequency shift of the emitted photons $\Delta\nu_{m_i,m_j}$ can be calculated for a transition between an upper $m_i$ and lower magnetic sub-level $m_j$ \cite{degl2014atomic}:
\begin{equation}\label{eq:shift}
    \Delta\nu_{m_i,m_j}=\left(g_{J_i}\times m_{i}-g_{J_j}\times m_j\right)\frac{\mu_0|B|}{h}
\end{equation}
Where $h$ describes the Planck constant, $\mu_0$ the Bohr magneton and $|B|$ the external magnetic field strength.
The frequency shift is relative to the unsplitted line-center frequency $\nu_0$ of the transition $i\rightarrow j$. The frequency $\nu_{m_i,m_j}$ can be described as follow:
\begin{equation} \label{eq:centerfrequency}
    \nu_{m_i,m_j}=\nu_0+\Delta \nu_{m_i,m_j}
\end{equation}
As a result, under strong magnetic fields ($B>0.2$ T\footnote[1]{The quoted field strength is the minimum value for the 801.48 nm and 842.47 nm emission line and depends in general on the Land\'{e} factors $g_{J_{i,j}}$.}), the plasma emission lines would exhibit strong effective broadening in the order of few GHz which is well above the thermal component ($\rm \approx 0.7~GHz$) resulting in an almost complete separation of magnetic sub-transitions.
 
\subsection{Optical emission}

The light emitted from an excited level $i$ to a lower level $j$ is proportional to the upper state density and transition probability expressed trough the Einstein coefficient. However, in optically thick plasma conditions, the emitted light will be reabsorbed by the lower state while propagating through the plasma on its way to the detector. Therefore, the intensity of a line emission reaching the detector $I_{i,j}$ can be expressed as \cite{schulze2008robust}
\begin{equation}
  I_{i,j}=\Theta n_i\gamma(n_j)A_{i,j}
\end{equation}
The measured intensity compared to the optically-thin plasma is reduced by the factor $\gamma(n_j)$, which accounts for self-absorption. The self-absorption depends on the absorption strength, lower state density and propagation length.
The factor $A_{i,j}$ is the Einstein coefficient for spontaneous decay, $n_i$ the upper state density and the factor $\Theta$ accounts for optical geometry and sensitivity of the spectrometer. 

\begin{figure}[h]
	\centering
	\includegraphics[width=0.5\linewidth]{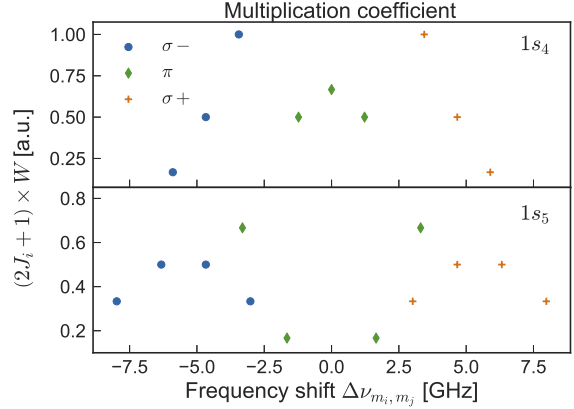}
	\caption{Calculated branching factors by (\ref{eq:Einstein}) represented with corresponding frequency shift calculated for magnetic field strength of B=0.3 T. Upper figure corresponds to 842.47 nm and lower for 801.48 nm transition in argon.}
	\label{fig:Einstein}
\end{figure}

The description of the intensity for an optically thick plasma holds also for the case B $\neq$ 0 where due to the splitting of the energy levels the total intensity is then described as a sum over all allowed transitions between upper and lower magnetic sub-levels categorized in $\sigma -$, $\pi$ and $\sigma +$ transitions (figure \ref{fig:schema-split}). They represent the transitions where $\Delta m_{j,i}=-1,0,1$ respectively and describe the right hand circular polarized, linear polarized and left hand circular polarized light. The total intensity emitted in a $\rm 2p\rightarrow \rm 1s$ transition can be summed up as $I_{\rm total}=I_{\pi}+I_{\sigma-}+I_{\sigma+}=I_{\pi}+2I_{\sigma}$ since $I_{\sigma+}=I_{\sigma-}=I_{\sigma}$. 
It is important to mention that linear polarized light is a superposition of two opposite circular polarized lights. This has an impact on the reconstruction of absorption profiles which will be discussed in more detail in section \ref{sec:resultsanddiscussion}. 
Each sub-transition can be described by the magnetic sub-level density $n_{m_i}$ of the upper state, the self-absorption factor which now depends on the magnetic sub-level density of the lower state $\gamma(n_{m_j})$ and the Einstein coefficient $A_{J_{i},J_{j},m_{i},m_{j}}$ between magnetic sub-level states $m_i$ to $m_j$. The formalism to calculate the Einstein coefficient between magnetic sub-levels can be written as \cite{takacs1996polarization,jacobs1999angular}:
\begin{equation}
\begin{split}
A_{J_i,J_j,m_i,m_j}= \left(2J_i+1\right)A_{ij}\times W\\\
     W=\left|\left(\begin{matrix}
			    J_j & q & J_i\\
                               -m_j & \Delta m_{j,i} & m_i\\
 \end{matrix}\right)\right|^2
\end{split}
\label{eq:Einstein} 
\end{equation}
The expression W is the Wigner 3-j coefficient to the power of two with the multi-polarity of the transition $q$ equivalent to 1 for electrical dipole transitions, and difference in the magnetic quantum number of the transition $\Delta m_{j,i}= m_j-m_i$. $A_{ij}$ describes the Einstein coefficient for spontaneous decay of the observed transition in unmagnetized case. Figure \ref{fig:Einstein} shows the calculated branching factors $(2J_i+1)\times W$ of the Einstein coefficient describing each sub-transition between magnetic sub-levels of $J_i=2\rightarrow J_j=1$ (top figure) and $J_i=2\rightarrow J_j=2$ (bottom figure) states, that can be further used in interpretation of measured $\rm 1s_4\rightarrow \rm 2p_8$ and $\rm 1s_5\rightarrow \rm 2p_8$ absorption profiles. Calculated factors are represented including corresponding frequency shift for each sub-transition calculated for magnetic field strength of $B = 0.3$ T.   

While the absorption strength is dependent on the electron configuration by quantum numbers $J_{i,j}$ and $m_{i,j}$ the resulting frequency shift of each sub-transition will be defined by the magnetic field strength.  

Calculated transition probability for each sub-transition will further define the intensity of emission lines originating from each upper magnetic sub-level and also define the self-absorption. 

 An analytic formula introduced by R. Mewe \cite{mewe1967relative} for the self-absorption coefficient is used to calculate each lower magnetic sub-level self-absorption coefficient $\gamma(n_m{_j})$ .

\begin{equation}
    \gamma(n_m{_j})=\frac{2-\exp(-10^{-3}\kappa(\Delta \nu=0)\times L)}{1+\kappa(\Delta \nu=0)\times L}
    \label{eq:selfcoeff}
\end{equation}
The expression $\kappa(\Delta \nu=0)$ describes the absorption coefficient at the line center of an individual absorption profile, for a given set of magnetic quantum numbers. $\Delta \nu$ equals $\nu_{m_i,m_j}$ from (\ref{eq:centerfrequency}) which describes the line center position and $L$ the characteristic length, which is dependent on the discharge geometry. For a cylindrical discharge geometry the characteristic length is assumed to be the radius of the discharge tube like proposed by R. Mewe.

The absorption coefficient at the line center for a magnetic sub-level transition is given by

\begin{equation}
    \kappa(\Delta \nu=0)=\frac{\lambda^3A_{J_iJ_j,m_im_j}}{8\pi }\left(\frac{M}{2\pi k_BT}\right)^{1/2}n_m{_j}
    \label{eq:abscoeff}
\end{equation}
where a mainly Doppler-broadened frequency distribution with temperature $T$ was assumed with the emission wavelength of the observed transition $\lambda$ and the mass $M$ of argon. The temperature was around 400 K on average measured from LAS and was used for the calculation.
 Atomic constants ($J_{i,j}$, $A_{i,j}$, $M$) and the Land\'{e} factors for calculations were
taken from NIST database \cite{NIST_ASD}.

\subsection{Tunable diode laser absorption spectroscopy}
\label{subsec:absorption}

Tunable diode laser absorption spectroscopy (TDLAS)
is a commonly-used technique to measure state densities based on the portion of the light absorbed in induced transition. The absorbed portion of the laser light is correlated to the number state density via the Beer-Lambert-Law \cite{sushkov2012time}:
\begin{equation}
    I=I_0\exp\left[-\kappa(\nu)l\right]
    \label{eq:lambert}
\end{equation}
The initial intensity $I_0$ is attenuated depending on
the optical path length $l$ in the absorbing medium and the
wavelength-dependent line integrated absorption coefficient $\kappa(\nu)$.
Recorded absorption structure was evaluated using Kapteyn package’s curve fitting routine \cite{terlouw2015kapteyn} where profile of each sub-transition was simulated with Gaussian (thermal) distribution function and superimposed to recreate the measurements. The magnetic sub-level density $n_m{_j}$ was calculated by fitting individual reconstructed absorption profiles from the level $m_j$ to a higher lying level $m_i$ to estimated the individual line integrated absorption coefficient $\kappa(\nu)$. The magnetic sub-level density can be calculated by \cite{kunze2009introduction}
\begin{equation}
    n_m{_j}=\frac{8\pi}{c^2}\frac{\nu_{m_i,m_j}^2}{A_{J_iJ_j,m_im_j}}\kappa(\nu)
\end{equation}
The degeneracy factors were not included since individual transitions between upper magnetic sub-levels and lower magnetic sub-levels are observed. The vacuum speed of light is expressed by $c$, the splitted central wavelength of the transition $\nu_{m_i,m_j}$ is given in (\ref{eq:centerfrequency}).

\section{Experimental Setup}

\begin{figure}[h]
	\centering
	\includegraphics[width=0.5\linewidth]{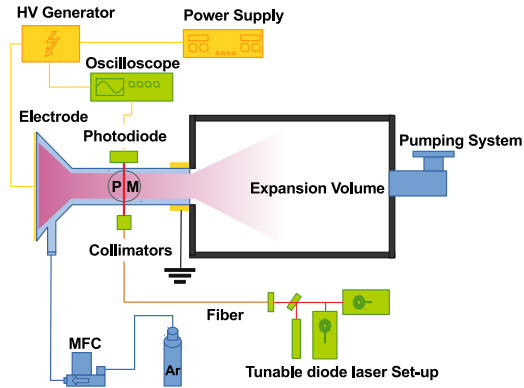}
	\caption{Schematic representation of the experimental setup.}
	\label{fig:set-up}
\end{figure}

The schematic of the experimental setup showing an overview of the used instrumental equipment and discharge configuration is presented in figure \ref{fig:set-up}.
Compact plasma jet is made out of glass and consists of a cone with 26 mm base diameter and height with a followed tube of 4 mm inner diameter extending from the top. Open end of the tube was attached to 500 mm long expansion chamber with the vacuum system attached at the end. The electrodes were made of aluminum tape attached from the outside of the jet on the cone base and at the end of the tube.

A base pressure of 4$ \times 10^{-2}$ Pa could be reached. For a gas flow of 32 sccm, the pressure inside the tube was estimated to be 100 Pa based on the axial TDLAS measurements and is described in more details in our previous study \cite{kaupe2018phase}. A 30 kHz sinusoidal high voltage ($V_{pp}\approx$ 4 kV) signal was used to drive the discharge. The external magnetic field was introduced by two cylindrical permanent magnets (6 mm in diameter), indicated as PM in figure \ref{fig:set-up}, inducing a magnetic field of 0.3 T. The magnets were mounted perpendicular to the line of sight at the opposite sides of the discharge tube.

Laser absorption measurements were done using two tunable diode lasers mounted on the same optical bench in a commonly used configuration, including argon reference cell and a Fabry-P\'{e}rot interferometer to monitor laser scanning range and quality. Absorption at 801.48 nm was targeted with \textit{TOPTICA DLC pro} while 842.47 nm transition was scanned by \textit{TOPTICA DLC 100} laser. A system of collimators and multimode $200~\mathrm{\mu m}$ optical fiber was used to manipulate the laser beams, producing an unpolarized probing laser light. Saturation effects were avoided by keeping the laser intensity below 5 $\mathrm{\mu W}~\mathrm{mm^{-2}}$ for all TDLAS measurements. Two band pass filters at 800 nm and 840 nm with a full width at half-maximum of 10 nm were used in order to suppress the rest of the plasma emission. The laser scan and plasma-driving high voltage signal
were synchronized by a trigger signal indicating the beginning of the voltage cycle (negative-to-positive voltage zero crossing) produced by the power supply. By simultaneously recording the voltage zero crossing, Fabry-P\'{e}rot and laser absorption signals it was possible to reconstruct the absorption profiles for different points within the voltage cycle resulting in phase resolved LAS.

The optical emission spectra was recorded by Intensified-CCD (PI-MAX4 1024f) attached to a 0.5 m monochromator in Czerny-Turner configuration. The measurements were done in accumulative mode, locking on the trigger output from the power supply achieving sub-microsecond time resolution.

The absorption and emission measurements were done using identical optics crossing the tube in radial direction in orientation perpendicular to the magnetic field lines as sketched in figure \ref{fig:set-up}. A linear polarization filter between the plasma and the collimator was used to isolate specific polarization from the unpolarized probing laser beam. With linear polarization filter oriented parallel to the magnetic field lines only $\pi$ transitions could be excited while perpendicular orientation would induce both $\sigma\pm$ transitions accounting that linear polarization can be expressed as superposition of left and right hand circular polarized light \cite{ida1998polarization}.

\section{Results and Discussion}
\label{sec:resultsanddiscussion}

\begin{figure}[h]
	\centering
	\includegraphics[width=0.5\linewidth]{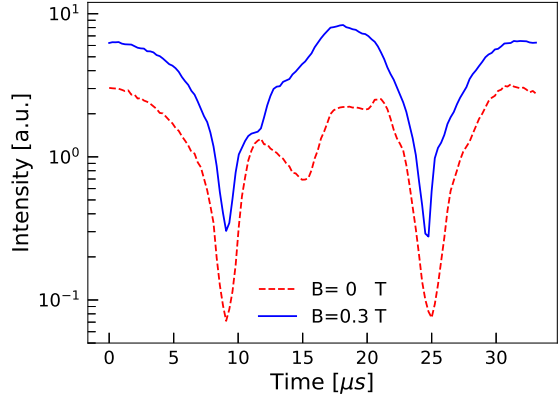}
	\caption{The phase-resolved intensity of 842 nm for 0 T and total intensity for 0.3 T recorded with an ICCD camera for 100 Pa. }
	\label{fig:842nm_intensity}
\end{figure}

Phase dependent development of plasma produced in same configuration and similar conditions has been investigated in previous works \cite{kaupe2018phase,kaupe2019phase} where strong variations in plasma parameters as well as in $\rm 1s$ state densities were observed. Plasma dynamics under the conditions investigated here illustrated by phase resolved emission of 842.47 nm argon line is shown in figure \ref{fig:842nm_intensity}. 
It is evident that presence of the magnetic field increases the intensity of the plasma emission within the whole cycle while also changing the axial distribution of light inside the jet. Nevertheless, strong phase dependent variations are still visible indicating strongly collisional and dark (recombinational) parts of the cycle. 
Such phase dependent variations in collisional-radiative dynamic would clearly induce density variations in targeted $\rm 1s$ states, however it is less clear if the density distribution among the magnetically separated levels would also be effected.

\begin{figure}[h]
	\centering
	\includegraphics[width=0.5\linewidth]{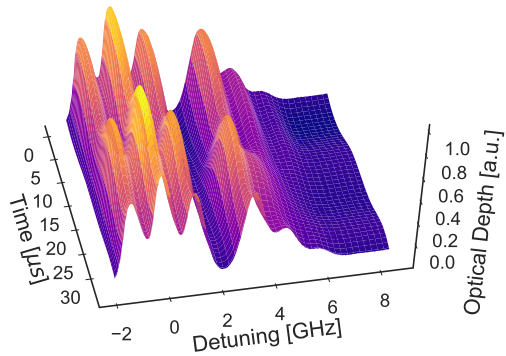}
	\caption{Phase-resolved absorption profile structure for $\rm 1s_4$. }
	\label{fig:1s4_3d}
\end{figure}

With the presented tunable laser absorption configuration, the phase-resolved absorption profiles were successfully reconstructed for targeted magnetic sub-level transitions of $\rm 1s_4$ and $\rm 1s_5$ states. An example of the phase-resolved absorption of the allowed $\pi$ and $\sigma +$ transitions of the $\rm 1s_4\rightarrow 2p_8$ excitation are shown in figure \ref{fig:1s4_3d}. The splitting of the energy levels will result in broad absorption structure extending over 12 GHz demanding high stability of the laser and fast laser scan. Since the absorption structure is symmetrical around the unshifted line center it was enough to scan over only one (positively or negatively shifted) part of the profile. From figure \ref{fig:1s4_3d} it is clear that all absorption components exhibit systematic variations within the cycle while only careful reconstruction of each component can result in estimations of sub-level number densities. Also, under applied magnetic field of 0.3 T, profiles were separated enough to provide reliable reconstruction. 

\begin{figure}[h]
	\centering
	\includegraphics[width=0.5\linewidth]{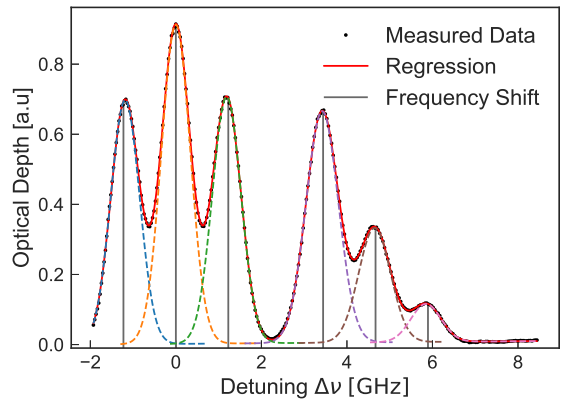}
	\caption{Measured absorption structure consisting of frequency shifted single sub-transitions of $\rm 1s_4\rightarrow \rm 2p_8$ transition. First three peaks (blue, orange, green) represent the $\pi$ transitions and the right side the $\sigma +$ transitions (purple, brown, pink). Dotted profiles represents Gaussian-profiles and vertical lines represent the theoretical calculated shifts. }
	\label{fig:1s4_profile}
\end{figure}

The typical example of reconstructed $\rm 1s_4\rightarrow \rm 2p_8$ absorption structure is presented in figure \ref{fig:1s4_profile} targeting a set of three $\pi$ and and three $\sigma +$ transitions as measured with unpolarized laser. 
The first three absorption profiles (from left to right) represent the $\pi$ polarized transitions while the following three absorption profiles are $\sigma +$ polarized corresponding to the transitions as indicated in figure \ref{fig:schema-split}. 
Both polarization structures consist of three transitions originating from different magnetic sub-levels of the $\rm 1s_4$ resonant state. The densities of magnetic sub-levels were reconstructed from the set of $\pi$ transitions due to a better separation of the absorption profiles and higher Einstein coefficients. 
The overlap of individual components of $\sigma \pm$ polarity is more pronounced causing a higher error in the reconstruction of the magnetic sub-level densities and positions of the line centers. 
The most shifted transition $m_i=0\rightarrow m_j=1$ from the $\sigma +$ (or $m_1=0\rightarrow m_j=-1$ from the $\sigma -$) has the smallest Einstein coefficient which consequences in smallest absorption. This causes complication in the reconstruction of the magnetic sub-level densities where the absorption is less pronounced especially in the dark parts of the discharge cycle. 
The measured structure was fitted with a set of 6 Gaussian profiles estimating their frequency shifts that could be further used in (\ref{eq:shift}) and (\ref{eq:centerfrequency}) to estimate the magnetic field strength. A good agreement with the theoretical calculated frequency shift assuming 0.3 T, indicated as grey vertical lines in figure \ref{fig:1s4_profile} and fitted line center was found. The intensities of the absorption profiles mostly dependent on the calculated Einstein branching factors from (\ref{eq:Einstein}) and were used to evaluate the magnetic sub-level densities.

Evaluation of the number densities from individually reconstructed absorption profiles would have to consider the fact that the measurements were done by unpolarized laser light. This way the $\rm \pi$ polarization is probed by the whole laser intensity while $\rm \sigma +$ transitions are only interacting with one half of the laser light having same circular polarization. This fact is reflected in measurements by equal intensities of the first profile of $\rm \pi$ and first profile of $\rm \sigma +$ sub-transitions that are originating from the same magnetic sub-level ($m_j=1$) of $\rm 1s_4$ while having different transition probabilities (see figure \ref{fig:Einstein} (top) where transition probability of the $\rm \sigma +$ line is twice higher than the $\rm \pi$ component). Neglecting the laser light polarity would result in different density estimation of the same lower level. Accounting the fact that the unpolarized light is superposition of both circularly polarized components \cite{ida1998polarization} would result in identical estimation of the density of the targeted states done from both components.
\begin{figure}[h]
	\centering
	\includegraphics[width=0.5\linewidth]{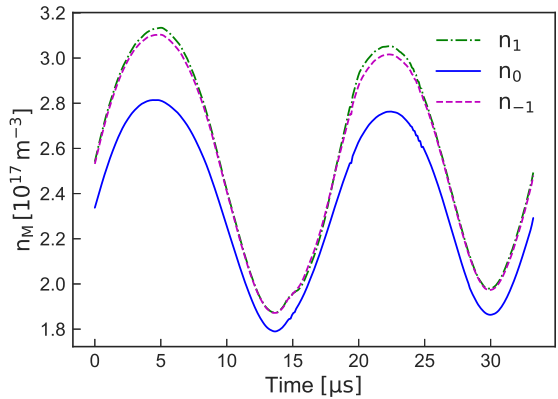}
	\caption{Reconstructed $\rm{1s_4}$ magnetic sub-level densities from $\pi$ transitions.}
	\label{fig:1s4_subdensities}
\end{figure}

The evaluation of $\rm 1s_4$ magnetic sub-level densities was based on modeling three $\pi$ components due to better reliability. Calculated phase dependent developments of $\rm 1s_4$ magnetic sub-levels are shown in figure \ref{fig:1s4_subdensities} where systematic variations in densities caused by modulation of voltage amplitude are clearly visible. Small but clearly visible population imbalance among the magnetic sub-levels is detected with less populated $m_j=0$ state creating positive alignment. Appearance of an alignment gives rise to linear polarization of the emission 
emitted from such set of magnetic sub-levels \cite{fujimoto2008plasma}. Sub-levels $m_j=\pm 1$ have almost identical densities within the whole cycle as expected due to the symmetry of the splitting and due to identical transition probabilities. 
The reconstruction from the $\sigma +$ components showed almost identical values for higher densities (high absorption at around 5 $\rm \mu s$ and 23 $\rm \mu s$) while considerable deteriorating in the rest of the cycle. The total $\rm 1s_4$ density under an external magnetic field was slightly higher than usually observed in our previous investigations without an external magnetic field. Such relatively high number densities could induce considerable self-absorption which will be evaluated latter in the text.

\begin{figure}[h]
	\centering
	\includegraphics[width=0.5\linewidth]{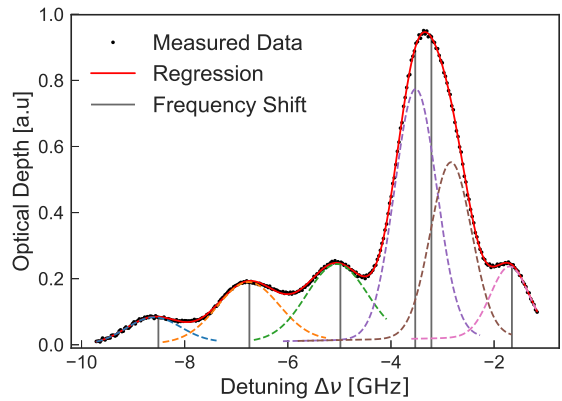}
	\caption{Measured absorption profile with predicted frequency shifts and single profile for $\rm{1s_5}$ transition. The blue, yellow, green and brown profiles represent the $\sigma -$ transitions the purple and pink respectively the $\pi$ transitions. Dotted profiles represents Gaussian-profiles and vertical lines represent the theoretical calculated shifts. }
	\label{fig:1s5_profile}
\end{figure}

Typical absorption structure targeting $\rm 1s_5$ metastable magnetic components at 801.48 nm is presented in the figure \ref{fig:1s5_profile}. The structure represents one half of the total magnetic absorption structure and consists of two $\pi$ and four $\sigma -$ transitions having negative frequency shift. The central line (without the shift) does not exist due to forbidden $m_i=0 \rightarrow m_j=0$ transition when $J_i=J_j$.   
In contrast to the previous case, the absorption components are not so clearly separated resulting in strong overlap and further causing difficulties in the reconstruction.

To overcome this issue a polarization filter was used to isolated laser component perpendicular to the magnetic field lines in order to probe only $\sigma -$ transitions. In this configuration only four $\sigma -$ components were detected that were further used in reconstruction of the $\rm 1s_5$ magnetic sub-level densities. It should be mentioned that $\pi$ absorption structure for the case where $J_i=J_j$ can not reconstruct all magnetic sub-level densities since the $m_i=0\rightarrow m_j=0$ is a forbidden optical transition and therefore not measurable. 

\begin{figure}[h]
	\centering
	\includegraphics[width=0.5\linewidth]{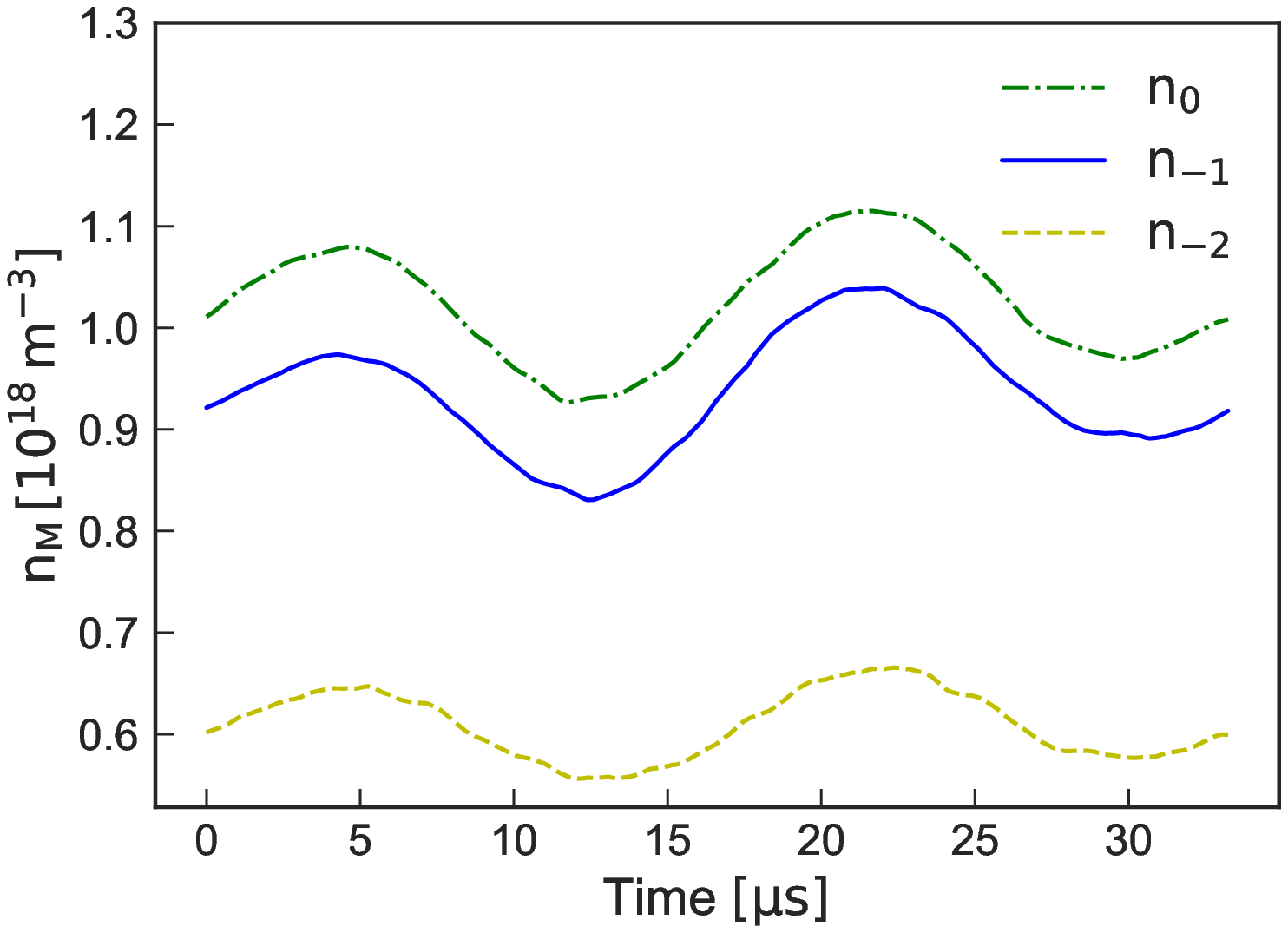}
	\caption{Reconstructed $\mathrm{1s_5}$ magnetic sub-level densities from $\sigma$+ transitions.}
	\label{fig:1s5_subdensities}
\end{figure}

The reconstructed phase dependent set of $\rm 1s_5$ magnetic sub-level densities is represented in figure \ref{fig:1s5_subdensities} for  $m_j=-2,-1,0$ states. Besides phase dependent variations, clearly visible population imbalance is detected presenting negative alignment with $n_{m_j=0}$ having the highest density. The population difference is stronger than in $\rm 1s_4$ sub-level states with almost 40 $\%$ less populated $m_j=2$ state.

 All magnetic sub-level densities from both states qualitatively resemble the trend observed for light emission from $\rm 2p_8$ state in figure \ref{fig:842nm_intensity}. The variation of the magnetic sub-level densities during the discharge cycle is more pronounced for the resonant then for the metastable state.

Reconstructed $\rm 1s_4$ and $\rm 1s_5$ magnetic sub-levels showed visible imbalance in density distribution, positive and negative alignment respectively, as a result of the collisional and radiative coupling primarily with ground and $\rm 2p_8$ excited states. Therefore a modeling of $\rm 1s$ state densities would, besides the radiative part, have to consider a less known electron impact excitation cross-section from the ground and to the $\rm 2p_8$ magnetic sub-levels. Although such a modeling would be highly advantageous in understanding of a magnetized plasma emission spectra, the task is highly non-trivial and out of the scope of this investigation.

Here, we will investigate the impact of splitting of the $\rm 1s$ states on plasma emission spectra focusing on description of light transport properties. For this purpose, the reconstructed magnetic sub-level state densities were further used to describe the self-absorption of individual emission components of the observed $\rm 2p_8\rightarrow \rm 1s_{4,5}$ magnetic transition structure. Densities of the states averaged over the cycle were used in further evaluations and comparison without a loss of generality. 
At first, each transition from the upper $\rm 2p_8$ state was described assuming equal density distribution among 5 $\rm 2p_8$ magnetic sub-levels and considering the appropriate branching Einstein coefficients. For each individual transition the self-absorption factor was calculated by (\ref{eq:selfcoeff}) and (\ref{eq:abscoeff}). The characteristic absorption length of 2 mm was used due to the high aspect ratio of the discharge geometry \cite{mewe1967relative} resulting in satisfactory estimations as shown in our previous work \cite{kaupe2018phase}.

\begin{figure}[h]
	\centering
	\includegraphics[width=0.5\linewidth]{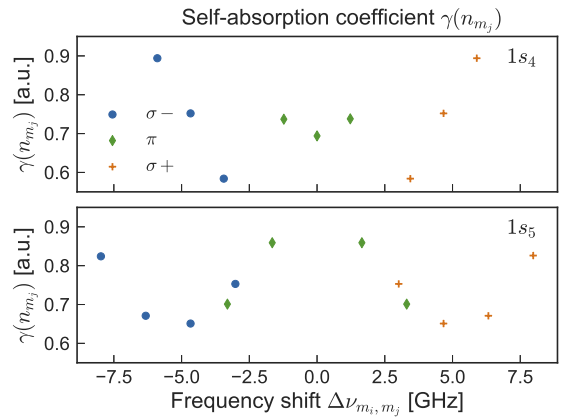}
	\caption{Calculated self-absorption coefficients for observed transitions between $\rm 2p_8$ magnetic sub-levels and $\rm 1s_4$ (top) and $\rm 1s_5$ (bottom) magnetic sub-levels.}
	\label{fig:self_absorption}
\end{figure}

The resulting transmission of each component in $\rm 2p_8\rightarrow \rm 1s_4$ and $\rm 2p_8\rightarrow \rm 1s_5$ transitions are presented in figure \ref{fig:self_absorption} up and down respectively. The transitions with low transition strength exhibit relatively small absorption of about 10 $\rm \%$ while stronger lines can be attenuated up to 40 $\rm \%$. Both transition sets are moderately absorbed due to relatively high $\rm 1s$ densities and high transition strength $\rm A_{842nm}=2.15\times 10^7$. The presented difference in the self-absorption of individual transitions can be further used to estimate the effects of self-absorption on the polarization of the emission lines. While equal light intensity is emitted in $\pi$ and $\sigma$ components from $\rm 2p_8$ magnetic sub-levels (assuming equal $\rm 2p_8$ magnetic sub-level densities, no alignment), uneven self-absorption could result in overall polarization of the light escaping the plasma. Considering alignment of $\rm 1s$ states observed in our measurements, the self-absorption would induce polarization of the emission lines with calculated averaged absorption of $\overline{\gamma_{842nm,\pi}}=0.723$, $\overline{\gamma_{842nm,\sigma}}=0.743$, and $\overline{\gamma_{801nm,\pi}}=0.780$, $\overline{\gamma_{801nm,\sigma}}=0.725$.

Total $\rm 1s$ densities calculated as a sum over all magnetic sub-levels, averaged over the voltage cycle, were estimated to be $n_{\rm 1s_4}= 7.44\times 10^{17}~m^{-3}$ and $n_{\rm 1s_5}=4.11\times 10^{18}~m^{-3}$. Assuming such $\rm 1s$ densities in unmagnetized conditions would result in increased self-absorption to relatively high $\gamma_{\rm 1s_4}=0.364$ and $\gamma_{\rm 1s_5}=0.317$ where larger part of light is lost on its way through the plasma.
As it is already clear, simple (non-magnetized) description of self-absorption in magnetized plasma would not work, so would not the branching factor method for estimation of $\rm 1s$ states densities based on self-absorption. Although this method is based on line ratios, we showed that the self-absorption can change due to the presence of an external magnetic field while their ratios could stay roughly the same. This would lead to misinterpretation of the measurements and underestimation of the densities.    

Presented comparison only roughly illustrates the difference in self-absorption mechanism induced by the presence of the magnetic filed. However, more rigorous and precise description would have to carefully reconsider the characteristic length $\rm L$ in (\ref{eq:selfcoeff}) which defines the intensity of the effect. In our experimental conditions, magnetic field had maximum strength between the magnets (where measurements were done) with slow decrease in the rest of the plasma resulting in inhomogeneous magnetic field. Therefore, the sub-transitions with maximum shift could experience excitation (by self-absorption) only from the light emitted from the plasma volume with highest magnetic fields. In the weaker fields (outside of the magnets), the shift would not be so strong so that the frequency of the light emitted from these positions would not match. Also, the less shifted transitions in the high magnetic field can be excited by the highly shifted transitions emitted from plasma in lower field. Such analysis would result in different characteristic length $\rm L$ of plasma that has to be evaluated for each transition depending on the in-homogeneity of the magnetic field distribution in plasma.

\section{Conclusion}

A method to treat line absorption measurements in plasma under the influence of an external magnetic field is provided resulting in reconstruction of the magnetic sub-level structure of $\rm 1s_{4}$ and $\rm 1s_{5}$ state of argon. The absorption structure was modeled accounting for Zeeman splitting of energetic levels and calculated branching of the Einstein coefficients for spontaneous decay. Optically active sub-transitions of different polarity were probed by unpolarized tunable diode laser light demanding special care in interpretation of the measurements. Reconstructed magnetic sub-level densities were further used in description of line emission properties focusing on self-absorption effect. 

Properties of 801.47 nm and 842.48 nm emission structure originating from $\rm 2p_8$ magnetic sub-levels were discussed by accounting for self-absorption effects on $\rm 1s_4$ and $\rm 1s_5$ magnetic sub-levels respectively. The presented analysis can be extended for any argon transition accounting for correct magnetic and total angular quantum numbers. 

The reconstructed magnetic sub-level densities of both states, $\rm 1s_4$ and $\rm 1s_5$, showed population imbalance, positive and negative, that can induce slight changes in the polarization of the emission lines escaping the plasma due to uneven self-absorption of polarized components. In general, the plasma becomes more transparent in the presence of an external magnetic field due to the splitting of the energy levels. Such effects would have to be carefully considered in description of emission spectroscopy in magnetized plasma. 
Criteria justifying presented analysis is proposed by R. Casini et al \cite{fujimoto2008plasma} demanding that the frequency shift is in the same range or higher than the thermal width, which is satisfied for $B>0.2$ T for observed transitions. Finally, the transport of plasma emission in inhomogeneous magnetic field was discussed, where difference in magnetic field strength would induce difference in the frequency shift, so that self-absorption would not work. Therefore, each transition between magnetic sub-levels would have to be described with individual effective absorption length based on the geometry of the discharge, magnetic inhomogeneity and details of the line splitting effect. 

The presented investigation demonstrates the approach in description and interpretation of the emission spectra of highly magnetized optically thick plasma conditions. While further modeling of magnetized plasma OES is highly non-trivial task challenged mostly by lack of data for collisional cross sections between sub-levels, here presented investigation could provide approach to correct interpretation and description of measured optical emission spectra. The discussed polarization dependent self-absorption could provide possibility to describe and model one more feature of a spectral line, which can be easily measured using standard polarization optics. 

\ack This work is supported by the Deutsche Forschungsgemeinschaft (DFG).

\section*{References}
\bibliographystyle{iopart-num}
\bibliography{paper.bib}

\end{document}